\input epsf
\magnification=1200
\def\Msun{M_\odot}
\def\go{\mathrel{\raise.3ex\hbox{$>$}\mkern-14mu
             \lower0.6ex\hbox{$\sim$}}}
\def\lo{\mathrel{\raise.3ex\hbox{$<$}\mkern-14mu
             \lower0.6ex\hbox{$\sim$}}}

\def\sec#1{\bigskip\noindent{\bf#1}\bigskip}
\def\subsec#1{\medskip\noindent{\bf#1}\medskip}
\newcount\secno
\newcount\eqnum
\def\equno{
\global\advance\eqnum by1\eqno({\the\secno.\the\eqnum})
}
\def\enam#1{
\xdef#1{(\the\secno.\the\eqnum)}
}
\centerline{\bf High-Resolution Simulations of Cluster Formation}
\vskip 1.0cm
\centerline{Tereasa G.\ Brainerd$^1$, David M.\ Goldberg$^2$ \&
Jens Verner Villumsen$^3$}
\vskip 1.0cm
\noindent
$^1${\it Boston University,
Dept.\ of Astronomy,
725 Commonwealth Ave.,
Boston, MA 02215}
\vskip 0.10cm
\noindent
$^2${\it Princeton University,
Dept.\ of Astrophysical Science,
Peyton Hall,
Princeton, NJ 08544}
\vskip 0.10cm
\noindent
$^3${\it Max-Planck-Institut f\"{u}r Astrophysik,
Karl-Schwarzschild-Strasse 1,
85740 Garching bei}

\noindent
{\it M\"{u}nchen,
Germany}
\vskip 2.0 cm
\centerline{ABSTRACT}
The formation history of rich clusters is investigated using a
hybrid N-body simulation in which high spatial and mass resolution
can be achieved self-consistently within a small region of a very large
volume.  The evolution of three massive clusters
is studied via mass accretion, spherically-averaged density profiles, 
three-dimensional
and projected shapes, and degree of substructure.  
Each cluster consists of $\sim 4 \times
10^5$ particles at the present epoch and in the case that rich cluster evolution
is well-described by a 1-parameter family, the simulations have 
sufficient resolution to demonstrate this.
At $z=0$ the clusters
have similar masses, $M(r \le 1.5h^{-1} {\rm Mpc})
\sim 2\times 10^{15} h^{-1} M_\odot$, and similar
spherically-averaged density profiles, however markedly different formation
histories are observed.
No single, dominant pattern
is apparent in the time variation of the mass accretion rate, the
cluster shape, or the degree of substructure.  Although not 
a statistically large sample, these objects suggest that the detailed
formation history of rich clusters cannot be characterized by a 
simple 1-parameter family.  
These results suggest that the use of
observations of
rich clusters over a wide range of redshifts to constrain cosmological
parameters may not be entirely straightforward.

\vskip 0.5cm
\noindent
{\it Subject headings:} cosmology: dark matter ---
cosmology: large-scale structure of the universe ---
cosmology: theory --- methods: numerical

\vskip 3.0cm
\line{\hfil Submitted to {\it The Astrophysical Journal}, June 1997 }

\vfil\eject

\secno=1
\eqnum=0
\sec{1. INTRODUCTION}

The evolution history of 
clusters of galaxies is a potentially powerful constraint
on theories of the formation of large-scale structure in the universe.  
Although a statistically large sample does not yet exist, it is
possible to identify clusters to high redshift (eg.\
Smail \& Dickinson 1995; Bower \& Smail 1997; Deltorn et al.\ 1997;
Luppino \& Kaiser 1997;
Steidel et al.\ 1997)
and it is hoped that eventually a complete description of the time
evolution of these objects will be obtained.  Additionally, detailed 
studies of clusters have yielded evidence for significant amounts
of substructure in many clusters, even those which appear smooth and
round in projection 
(eg.\ Beers \& Geller, 1983; Jones \& Forman 1984; Dressler
 \& Shectman 1988; West \& Bothun 1990;
Davis \& Mushotzky 1993; Miyaji et al.\ 1993; 
Mushotzky 1993; White, Briel \& Henry 1993; Bird 1994a,b; 
Zabludoff \& Zaritsky 1995).  Optical, X-ray, and
kinematic evidence  has been found for recent mergers of a number
of clusters with smaller systems.  It appears that at least one third
of all clusters are not fully relaxed systems and it is possible that
many clusters are still in the process of assembling even today. 
The fraction of clusters containing significant amounts of substructure 
at the present epoch is 
a potentially powerful constraint on the value of the density
parameter (eg.\ Richstone, Loeb \& Turner 1992; Bartelmann, Ehlers \&
Schneider 1993; Kauffmann \& White 1993;
Lacey \& Cole 1993; Mohr et al.\ 1995) and, therefore, cluster substructure
investigations are of considerable interest.

It has been hoped that
numerical simulations of cluster formation
would provide useful constraints on large-scale structure theories
via comparisons of the
formation histories of simulated clusters and the observed cluster
population.  Direct comparisons are, however,
problematical for a number of reasons.  First, pure N-body simulations
follow only the evolution of the dominant, dissipationless mass component
of the universe, neglecting hydrodynamics.  In such simulations 
a direct comparison of a theoretical distribution of
light (eg.\ galaxies and X-ray gas) to that of observed 
clusters is not possible.  Under the assumption that density peaks of
an appropriate mass scale correspond to the likely sites of galaxy formation,
however, it is possible to locate groups of
particles within the simulations that
may be associated fairly with the dark matter halos
of individual galaxies.  Additionally, from studies of 
the coherent weak shear field
associated with gravitational lens clusters, the mass of the clusters is
certainly dominated by dark matter and in addition it appears that the mass 
distribution within the clusters traces the smoothed light distribution
quite well (eg.\ Bonnet, Mellier \& Fort 1994; 
Fahlman et al.\ 1994; Smail et al.\ 1995;
Kneib et al.\ 1996; Seitz et al.\ 1996; Squires et al.\ 1996ab; Smail
et al.\ 1997).
Therefore, cluster simulations which include 
only a dark matter component should yield fairly reasonable results for
comparison with observation, at least to a good first approximation.

The worst problem to plague simulations of cluster formation
is simply one of resolution, both in terms of the gravitational
force calculation on small scales as well as the mass
per particle.  That is, within the cluster environment itself,
one would like to resolve the physical
scales associated with galaxies (distances of order a few tens of
kiloparsecs, using a mass per particle of order $10^9 \Msun$).
Ideally, of course, one would like to
achieve such resolution inside a simulation volume which is itself a
``fair sample'' of the universe (of order $10^7 h^{-3} {\rm Mpc}^3$).
Current computing platforms, however, do not allow such a high level of 
resolution throughout a large simulation volume.  

Therefore, high-resolution simulations of cluster formation often
follow a scenario in which a simulation of a large volume of the
universe is first run at moderate resolution.  The final timestep of this
simulation is then searched for peaks in the smoothed mass density which
would correspond fairly to clusters of galaxies.  A sphere of a given
radius (typically $r \sim 10h^{-1}$~Mpc) centered on the
peak is then excised from the initial conditions of the simulation and
populated with a large number of particles of small (sub-galactic) mass
(eg.\ Bromley et al., 1995; Carlberg, 1994).
These smaller peak particles are then evolved from the initial timestep to 
the present epoch, subject to an external potential which is intended
to model the correct tidal field due to the local universe that the
cluster would experience as it evolves.
Difficulties with such simulations are insuring: (1)
the radius of the sphere is large enough to include
all the mass that should be accreted by the cluster up to the present
epoch and (2) the model external potential fairly
represents the actual tidal field the cluster would experience
if one simply ran the entire simulation at an unachieveably high
resolution level.

Here we simulate the formation of 3 rich clusters at very high
resolution and investigate
the similarities and differences of their evolution histories.
All three clusters have similar masses at the present day
and for simplicity a standard cold dark matter universe is adopted.
The clusters form inside a large computational volume ($8.0\times 10^6
h^{-1} {\rm Mpc}^3$) and high resolution is achieved without the
use of either constrained initial conditions or the excision of peaks from 
a large-scale density field.  Instead, a hybrid N-body code
utilized.  This particular code allows high spatial and mass resolution to be
achieved simultaneously within small selected regions of a very large
primary simulation volume.  High resolution is obtained
by nesting small simulations self-consistently inside larger
simulations, resulting in a ``power zoom'' effect.
The basic premise behind the N-body code
used for the investigation is outlined in \S2.  Details of the simulations
performed are summarized in \S3, results of the analysis of the
simulations are presented in \S4, and a discussion of the results
is given in \S5.  

\secno=2
\eqnum=0
\sec{2. HIERARCHICAL PARTICLE MESH (HPM)}

The code used to run the simulations is the Hierarchical Particle
Mesh (HPM) code written by J.\ V.\ Villumsen (Villumsen, 1989).  The
heart of the HPM code is a standard particle mesh (PM) cosmological
simulation in which mass density is assigned to a grid using a 
cloud-in-cell (CIC) weighting scheme and Poisson's equation is solved
using fast Fourier techniques.  Although very fast, PM simulations suffer
from limited spatial resolution, the force being softer than
Newtonian on scales smaller than about 2 grid cells.  In order to
gain both spatial and mass resolution in a small region of 
the primary simulation volume, the HPM code allows
small simulations (``subgrids'') to be nested self-consistently within 
the main simulation.  
By nesting subgrids inside subgrids one can progressively
build up to very high resolution in a limited region of the total
simulation volume.  Details of the force calculations and the generation
of initial conditions for multi-grid simulations are given in 
Villumsen (1989); here we present only a brief outline of the premise
behind HPM.

It is important to note that
a multi-grid simulation using HPM is an {\it iterative} process.  To begin,
an ordinary PM simulation of a large volume of the universe is 
run from the desired starting epoch ($z\sim 30$) to the present
epoch.  Periodic boundary conditions are imposed on this grid and the
simulation is carried out in a manner similar to all conventional PM
simulations of the formation of large-scale structure.  Throughout, 
this large grid shall be referred to as the ``top grid''; it constitutes
the fundamental computational volume of the simulation.  

A small region of interest which is to be run in high resolution mode
(eg.\ a cluster environment) is then selected from the {\it final timestep} of
the top grid.  Using the previously recorded timesteps
of the top grid calculation, those top grid particles which
pass through the region of interest (plus a generous buffer zone)
at any time during the course of the simulation are tagged.  
The entire simulation is then reeled back to the uniform
grid stage and for each of the top grid particles that were tagged
as having passed through the region of interest, a set
of smaller particles is generated for the subgrid calculation.
This is done in the following manner.
Each of the tagged top grid particles defines a cubical box of length $l_t$,
equal to the interparticle spacing in the top grid.  Allowing the
subgrid to be a factor of $f$ smaller than the top grid, a virtual grid
of subgrid particles is then generated with a spacing of $l_s = l_t/f$
and any virtual particle in a box defined by a top grid particle is
counted as a subgrid particle.
At this point the subgrid particles constitute a uniform grid which
is fully sampled inside the subgrid volume and only partially sampled
outside it.  Initial conditions for {\it both} the top grid and the subgrid
are then generated by perturbing the top grid and subgrid particles
away from their respective uniform grids.  (Seeds identical to the
seeds used to generate the first set of top grid initial conditions 
are used so that the initial conditions for the top grid in the
multi-grid calculation will be identical to the initial conditions
for the top grid alone.)

The full multi-grid calculation is then run with the two sets
of initial conditions, the top grid and the subgrid being evolved forward
in time simultaneously.  
The important points to note are: (1) there is no ``back-reaction''
from the subgrid to the top grid (i.e.\ the top grid simulation
runs completely unaware of the subgrid simulation), 
(2) the potential in
the subgrid is computed using both the small particles in the
subgrid and the potential from the top grid (i.e.\ the force field
from the top grid acts as an external field on the subgrid
simulation), and (3) unlike the top
grid, the subgrid utilizes isolated boundary conditions so that
subgrid particles may enter and exit the subgrid region over the
course of the simulation.
Additionally, the subgrids may either be kept stationary
throughout the course of the simulation or they may be allowed to
move (eg.\ to follow the growth of an object which has a large
streaming velocity).

An HPM simulation is constrained to use the same number of grid
cells in both the top grid and subgrid simulations.  Therefore 
by using a subgrid which is factor of $f$ smaller than the top
grid, the gain in spatial resolution in the subgrid region is necessarily 
a factor
of $f$.  The total number of particles used in the subgrid may, however,
vary from that used in the top grid.  Allowing an identical number
of top grid and subgrid particles across a uniform grid, a subgrid
which is a factor of $f$ smaller than the top grid results in a
mass per particle that is a factor of $f^3$ smaller than in the top
grid.  However, in a high density region of the simulation (eg.\ a
cluster) it is oftentimes necessary to reduce the number of uniform
grid particles in the subgrid compared to that of the top grid in order
to remain within the available machine memory.

Very high resolution in an HPM simulation may be obtained by further
nesting subgrids within subgrids.  Again, this is an iterative process.
At the end of a two-level (top grid plus one subgrid) calculation,
a region of interest is identified within the subgrid.  All of the subgrid
particles which pass through that region of interest over the course
of the simulation are tagged and the entire simulation (top grid plus
subgrid) is reeled back to the uniform grid stage whereupon a second
subgrid is generated within the first subgrid utilizing 
all the tagged particles
from subgrid \#1.
Using the same random number generator seeds as were used previously,
initial conditions for each of the grids (top grid plus two subgrids)
are generated by perturbing each of the three sets of particles away
from their respective uniform grids.  Again, in the multi-grid
calculation the grids are evolved
forward in time simultaneously, there is no back-reaction from
``parent'' grid to ``child'' grid, and the force field from the 
``parent'' grid acts as an external field for the force calculation
in the ``child'' grid.

Since the ``child'' particles experience high frequency power in
their subgrid that their ``parent'' particles do not, the ``child'' particles
will not move exactly in concert with their ``parent'' particles.  
However, the ``child'' particles do not stray very far from the general
location of the ``parent'' since the extra high frequency power does
not induce large streaming velocities.  Should a ``parent'' particle
exit/enter the region of a subgrid, it takes its ``children'' out of/into
the region in a smooth manner.  A simple consistency check involves comparing
the number of ``child'' particles found inside a given subgrid at a
particular time to the number of ``parent'' particles also within the subgrid
at the same time.  The ratio of these numbers should be of order 
the cube root of the ratio of the particle masses in the two grids (but
will not be exactly equal to this value
owing to the smooth manner in which the
``child'' particles enter and exit the subgrids).
This is, indeed, the case and for the simulations presented here
the ratio of the number
of ``child'' particles to ``parent'' particles differs from the
cube root of the ratio of the particle masses by an average of about 6\%.

Visual comparisons between the structures in ``parent'' and ``child''
grids show excellent agreement (see Fig.\ 1).  However, due to the higher force
resolution in the ``child'' grid, structures in the ``child'' grid
tend to be more concentrated than the analogous, more poorly resolved
structures in the ``parent'' grid.  That is, the force in the ``child''
grid is not as soft as in the ``parent'' grid, allowing structures to
collapse on smaller scales.

\secno=3
\eqnum=0
\sec{3. THE SIMULATIONS}

Three multi-grid simulations of the formation of clusters in a standard
CDM universe ($H_0=50$~km/s/Mpc,
$\Omega_0=1$,
$\Lambda=0$) were run.
All simulations consisted of 3-level calculations: 
a top grid of length $L_{\rm top}=200h^{-1}$~Mpc, inside which was nested a
subgrid of length $L_{\rm sub1}=33.3 h^{-1}$~Mpc, inside which was nested
a subgrid of length $L_{\rm sub2}=8.3 h^{-1}$~Mpc (comoving
lengths).  In all cases
$256^3$ grid cells were used, resulting in a grid cell length
of $32.6 h^{-1}$~kpc (comoving) in the smallest, highest resolution subgrid.  
A total of $128^3$ particles
were used in the top grid calculation, 
resulting in a mass per particle in the
top grid of $1.06\times 10^{12}h^{-1}\Msun$.  Owing to
machine memory limitations, the uniform subgrids were constrained to 
fewer particles than the top grid.  The particle
mass in the low-resolution subgrids (subgrid \#1)
was $3.90 \times 10^{10} h^{-1} \Msun$ while
in the high-resolution 
subgrids (subgrid \#2) it was $4.88 \times 10^{9} h^{-1} \Msun$.
At the end of the simulation there were $8.3 \times 10^5$ particles inside
the high-resolution subgrid containing cluster 1, corresponding to a
density of 179 particles per cubic megaparsec.  For cluster 2, there
were $7.3 \times 10^5$ particles inside its high-resolution subgrid at
the end of the simulation, corresponding to a density of 159 particles
per cubic megaparsec.  For cluster 3, there were $9.8\times 10^5$ 
particles inside its high-resolution subgrid at the end of the simulation,
corresponding to a density of 211 particles per cubic megaparsec.

The locations of the subgrids were specified as follows. To begin, a top grid
simulation was evolved from $\sigma_8 = 0.033$ to $\sigma_8 = 1.0$, where
$$ \sigma_8 \equiv \left< \left[ {\delta\rho \over \rho}(8h^{-1} {\rm Mpc})
\right]^2 \right>^{1\over 2} \eqno (1) $$
Identifying the final timestep of the top grid simulation as the
present epoch (redshift, $z$, of 0), the simulation began at $z=29$.  This
is a model which is somewhat under-normalized compared to the
COBE observations (eg.\ Bunn \& White 1997) and over-normalized
compared to the abundance of rich clusters (eg.\ Bahcall \& Cen 1993;
White, Efstathiou \& Frenk 1993; Eke, Cole \& Frenk 1996;
Viana \& Liddle 1996).
To determine the present-day locations of
rich clusters, the mass density field of the
top grid at $\sigma_8 = 1.0$ was smoothed with a Gaussian filter of length 
$1.5 h^{-1}$~Mpc and the locations of peaks determined.  From 
this smoothed density field the locations of three
of the largest density peaks were selected as the centers of
the first subgrids. 
For each of these subgrids, the top grid particles that
passed through the subgrid region (plus 20\% buffer zones) were
tagged, the simulations were reeled back to the uniform grid stage,
initial conditions for 2-level calculations were generated, and
the 2-grid simulations were then evolved from $\sigma_8 = 0.033$
to $\sigma_8 = 1.0$.  
The second, highest resolution
subgrids were then chosen to be centered on the centers of mass
of the clusters that formed in each of the first subgrids.
Again, using the timesteps of the 2-level calculations, the particles
in the first subgrids that passed through the regions of the second
subgrids (plus buffer zones) were tagged, the simulations were reeled
back to the uniform grid stage, initial conditions for 3-level
calculations were generated, and the 3-grid simulations were then evolved
from $\sigma_8 = 0.033$ to $\sigma_8 = 1.0$. 

The clusters investigated here correspond to the very largest (i.e.\ most
massive) of these objects that would typically form in 
CDM universes.  They do not, therefore, represent an
unbiased, ``average'' sample of clusters but may correspond fairly
with the ``richest'' clusters that would form in such
universes.  Tables 1, 2, and 3 contain summaries of various properties
of the clusters obtained from analyses of the highest resolution
subgrids.  At the end of the simulation all three clusters
have masses of order $2\times 10^{15} h^{-1} \Msun$ within the Abell
radius ($1.5h^{-1}$~Mpc), corresponding to of order $4\times 10^5$
particles in the highest resolution subgrids.

Shown in Fig.\ 1 are 
grey-scale pictures of the clusters at $\sigma_8 = 1.0$.
The level of grey indicates the logarithm of the
mass density along the
line of sight in the projection and each projection has
dimensions
$8.3 h^{-1} {\rm Mpc} \times 8.3 h^{-1} {\rm Mpc} \times 8.3 h^{-1}
{\rm Mpc}$.  That is, shown in Fig.\ 1 is a 2-dimensional compression
of a 3-dimensional 
volume corresponding to the full volume of the highest resolution
subgrid and each projection is centered on the center of mass
of the cluster.  The top panels show the clusters at highest resolution
(subgrid \#2) and the center panels show the clusters at
lower resolution (subgrid \#1).  The sizes of the pixels in the figure
correspond to the sizes of grid cells in the different subgrids and
reflect their relative levels of resolution.  In each case the high
mass density in the inner regions of the clusters results in a ``burned
out'' image, but in the outer regions of the clusters it is clear that
there are many smaller galaxy-sized mass concentrations.  (There are
also smaller galaxy-sized mass concentrations in the inner regions of
the cluster which are not visible in Fig.\ 1 due to the level of 
contrast.  See \S 4.6.)  The bottom panels in Fig.\ 1 show a comparison
of the mass density along the line of sight in the high- and low-resolution
subgrids.  Specifically, the grey-scale indicates the logarithm of
the ratio of the mass density in subgrid \#2 to the mass density in
subgrid \#1.  The comparison is done at the same (low) resolution as
subgrid \#1.  Overall the comparison is excellent (i.e. the image is fairly
flat at a moderate level of grey, indicating a density ratio of order
unity).  The largest discrepancy between the densities in the two subgrids
occurs near the ``edges'' of the clusters where the density in subgrid \#2
is less than in the corresponding regions of subgrid \#1
(i.e. white pixels).  The discrepancy is caused by the relative levels
of numerical softening in the two subgrids, the force being softer
on a larger scale
in subgrid \#1 than in subgrid \#2, resulting in less concentrated
structure in subgrid \#1 compared to subgrid \#2.  That is, in the outer
regions of the cluster the high-resolution version is somewhat less 
dense than the low-resolution version since, on the whole, the cluster is
more condensed in subgrid \#2 than it is in subgrid \#1.
Spherically-averaged
density profiles of the clusters computed using the two different subgrids
are, however, in excellent agreement at large radii (see Fig.\ 5).

\secno=4
\eqnum=0
\sec{4. RESULTS}

\subsec{4.1 Mass Accretion}

The growth of each of the clusters was investigated through: [1] the
mean infall distance of particles into the cluster, 
[2] the time evolution of the total mass of the cluster contained
within the Abell radius
and [3] the rate at which mass was
accreted within the Abell radius as a function of time.
In order to calculate each of these
quantities the center of mass of each cluster is required.
This was determined using the following iterative procedure.  
Starting with an initial
center of mass given by the location of the corresponding
peak in the smoothed mass
density field of the top grid, 
all subgrid particles within a radius of $3.0h^{-1} {\rm Mpc}$ 
were selected
and the center of mass of those particles was computed.  From this 
center of mass a new sphere of particles of radius $3.0h^{-1} {\rm Mpc}$
was selected and a new center of mass computed.  The process was 
repeated within a given subgrid until convergence was reached 
(of order 6 iterations).  Note that the centers of mass for each cluster 
computed independently
from the corresponding high- and low-resolution subgrids are
identical. Also, over the course of the simulations the 
centers of mass of the clusters have low streaming velocities
and they move total distances which are less than the mesh resolution
of the top grid.

Fig.\ 2 summarizes both the mean and maximum infall distances
of particles over the course of the simulations.
The points with error bars in Fig.\ 2 indicate the 
mean initial distance of subgrid \#1 particles from the centers of mass of
their respective clusters as a function of their distance from the
cluster centers of mass at the end of the simulation.  That is, the points
indicate the mean streaming distance of particles present in the clusters
at the end of the simulation as function of their distance from the
center of mass at the final timestep.  The error bars show one
standard deviation.  The open squares without error bars indicate the
{\it maximum} initial distance of any one particle from the center of
mass of its cluster versus its distance from that cluster at the end
of the simulation.
From this figure, then, over the course
of the simulations the mean distance traveled 
by particles found within the
Abell radii at the end of the simulations is of order
$12h^{-1} {\rm Mpc}$.  This is not at all surprizing since the mass of
the clusters within the Abell radius at $\sigma_8 = 1.0$ is of order
$2\times 10^{15} h^{-1} \Msun$, equal to the mass contained within
a uniform critical density sphere of radius $12h^{-1} {\rm Mpc}$.  The
maximum infall distance, however, is of order $18h^{-1} {\rm Mpc}$.
Thus, in order to follow all of the infall of mass into the
clusters over the course of the entire simulation,
a simulation that utilizes
either constrained initial conditions or the excision of peaks from 
a large-scale structure simulation would require a volume of 
$\sim 3\times 10^4 h^{-3} {\rm Mpc}^3$ to be simulated at comparably
high resolution. 
This is a factor of order 50 larger than the requisite volume for the 
highest resolution subgrid in the HPM calculation.

The masses of the clusters contained
within the Abell radius, $M(r = 1.5h^{-1}{\rm Mpc})$, 
and the rates at which mass was accreted
within the Abell radius, $\dot{M}(r = 1.5h^{-1}{\rm Mpc})$,
were computed as a function of lookback time
for each of the
three clusters.  The high-resolution subgrid particles
were used for these calculations and
the lookback time was computed by taking
$\sigma_8 = 1.0$ to be the present epoch.
Results for the evolution of the total amount of
mass within the Abell radius, normalized by the present-day
mass of the cluster within the same (comoving) distance,
are shown in Fig.\ 3.  It is clear from this figure that the
details of the evolution of the clusters are somewhat different in
each case but that all three gained of order 50\% of their present-day
mass within the past 5 to 8 Gyr .  The details of the {\it rate} at which
mass was accreted within the Abell radius
are shown in Fig.\ 4.  Here the rate at which mass was accreted by 
each cluster, 
$\dot{M}$, is shown as a function of lookback time and redshift, 
normalized by $\left< \dot{M} \right>$, the average rate at which the cluster 
accreted mass between $z=2$ and $z=0$.
From this figure it is clear that, although all three clusters have
similar masses at the present, no single pattern of mass accretion
dominates in the formation of the clusters.  Cluster 1 shows a monotonic
increase in mass accretion rate from $z=2$ to $z=0.2$, after which it
accretes virtually no mass.  Cluster 2, however, shows a monotonic decrease in
the mass accretion rate from $z=2$ to the present and cluster 3 
forms via a mass accretion rate which is roughly constant.

\subsec{4.3 Density Profiles}

Spherically-averaged differential density profiles, $\rho(r)$, are shown
in Fig.\ 5 for each of the clusters at the end of the simulations.  Squares
indicate the density profiles obtained using the 
high-resolution subgrid particles and triangles indicate the density profiles
obtained using the low-resolution subgrid particles.  Due
to the numerical softening of the force on small scales
the density is computed only on scales larger than two grid cells
($65h^{-1}$~kpc in the high-resolution subgrids and $260h^{-1}$~kpc
in the low-resolution subgrids).  Over
the length scales for which the density can be computed in both the
high- and low-resolution subgrids there is excellent agreement between
the two calculations and the small-scale density profile computed from the
high-resolution subgrid is clearly a smooth continuation of the larger-scale
density profile computed from the low-resolution subgrid.  
Also, but for a suggestion of a flattening in
the density profile of cluster 1, there is no turnover in the density
profiles at small radii, which is as expected in purely dissipationless
simulations.  The apparent flattening in the density profile of cluster 1
may be partially numerical in origin as it occurs on scales corresponding
to 3 grid cells and less.  A higher resolution simulation (eg.\ a third
subgrid) would be required to determine whether the trend is indeed real or 
purely an artifact.  A comparison of these density profiles
and corresponding density profiles obtained from
the particles in the top grid calculation 
is not warranted owing to the extremely
poor resolution of the clusters in the top grid (of order 1000 particles
in total
and a force softer than Newtonian on scales smaller than the Abell radius).

It is clear from Fig.\ 5 that the density profiles of the clusters
are not well-fit by a single power law over all scales.  This is to
be expected since it has been shown previously that CDM halos are well-described
by density profiles in which the logarithmic slope varies gently
(eg.\ Dubinski \& Carlberg 1991; Navarro, Frenk \& White 1995, 1996ab;
Cole \& Lacey 1996; and Tormen, Bouchet
\& White 1997).  On scales $r \le r_{200}$, where $r_{200}$ is the
radius inside which the mean interior overdensity is 200, 
a good two-parameter fit to the density profile is given by
$${\rho(r) \over \rho_c} = {\delta_c \over x(1+x)^2} \eqno (2)$$
(eg.\ Navarro, Frenk, \& White 1996b)
where $\rho_c$ is the critical density for closure of the 
universe, $x \equiv r/r_s$, and $r_s$ is a scale radius.  Here
$\delta_c$ is a dimensionless characteristic density.  By defining
the ``concentration'' of a halo to be $c \equiv r_{200}/r_s$, the
two-parameter fit above can be reduced to a one-parameter fit through
$$\delta_c = 
{200 \over 3}{c^3 \over \left[ {\rm ln}(1+c) - c/(1+c) \right] } \eqno (3)$$
(eg.\ Navarro, Frenk \& White 1996b).

Using the particles in the highest resolution subgrids, the values of
$r_{200}$ (the ``virial radius'') for the clusters were determined
at $\sigma_8 = 0.67, 0.83, 1.0$.  The virial radii evolve relatively little
from $\sigma_8 = 0.67$ to $\sigma_8 = 1.0$ and are of order $2h^{-1}$~Mpc
for each of the clusters (see Tables 1, 2, and 3 for specific values).
Again using the particles in the highest resolution subgrids, the variation
of the cluster overdensities with radius were evaluated on scales less than
$r_{200}$ and results are shown by the points in Fig.\ 6.  The solid
lines in this figure illustrate the best-fitting density profiles of
the form of equation (3) above. The values of the corresponding
scale radii, $r_s$, are given in each of the panels of the figure.  
But for a slight downturn in the
small-scale density profiles of cluster 2 at $\sigma_8 = 0.67$ 
and cluster 3 at
$\sigma_8 = 0.83$, there is very good agreement between the simulated
clusters and equation (3).  
Again, it is possible that the small-scale downturn
is numerical in origin.  The scaled density profiles in Fig.\ 6 are
all fairly similar and in the case of clusters 1 and 2 the scale radius
of the best-fitting profile evolves little from $\sigma_8 = 0.67$ to
$\sigma_8 = 1.0$; however, for cluster 3 the value of
$r_s$ changes appreciably
(by a factor of order 2) over the same time period.

\subsec{4.4 3-d Shapes \& 2-d Ellipticity Distributions}

The evolution of the 3-dimensional and projected shapes of the clusters
were computed at $\sigma_8 = 0.67$, 0.83, and 1.0 using the particles
in the highest resolution subgrids.  Since there is no hard ``edge''
to the clusters in terms of distinguishing those particles which are inside
the cluster and those which are not, we shall define the boundaries of the
clusters to be the virial radii ($r_{200}$) for the following analyses.

Using all particles within $r_{200}$ of the cluster centers of mass, the
3-dimensional cluster shapes were determined from a
standard moment of inertia analysis that yielded
the axis ratios $b/a$ and $c/a$ for each of the clusters (we define
$a > b > c$).  From the axial ratios a triaxiality parameter was
computed for each of the clusters:
$$T = {a^2 - b^2 \over a^2 - c^2} \eqno (4)$$
where $T = 0$ indicates a purely oblate object and $T = 1$ indicates a
purely prolate object.  We shall refer to objects  with 
$0 < T < 1/3$ as being nearly oblate, those with $2/3 < T < 1$ as
nearly prolate, and those with $1/3 < T < 2/3$ as triaxial.
Values of the cluster axial ratios and triaxiality
parameters are listed in Tables 1, 2, and 3 for $\sigma_8$ between 
0.67 and 1.0. From these tables it is clear that the evolution of the
shapes of the clusters are quite different in each case.  Although
cluster 1 and cluster 2 are both nearly oblate at the end of the simulation,
cluster 1 evolves from being triaxial at
$\sigma_8 = 0.67$ to being nearly oblate at $\sigma_8 = 1.0$ whereas
the shape of 
cluster 2 changes little over the same period of time and, as a result,
is always nearly oblate.
Cluster 3, on the other hand, is nearly prolate at the
end of the simulation but was nearly oblate at $\sigma_8 = 0.67$.

A more useful quantity for comparison of the evolution of the shapes
of simulated clusters to observed clusters is the ellipticity projected
on the plane of the sky.  In the case of the simulations the projected
ellipticity of the mass is the only
quantity which can be computed reliably (i.e.\ without having to resort
to assumptions about the degree to which mass would trace light).
Given recent advances in gravitational
lensing analyses of observed clusters, however, this seems a
reasonable quantity to compute.  That is, from analyses of the coherent weak
distortion of the shapes of background galaxies due to an intervening
gravitational lens cluster
it is possible to constrain the ellipticity of the projected
mass of clusters and, additionally, it is becoming clear that the smoothed
light distribution of clusters traces the mass quite well
(eg.\ Bonnet, Mellier \& Fort 1994; 
Fahlman et al.\ 1994; Smail et al.\ 1995;
Kneib et al.\ 1996; Seitz et al.\ 1996; Squires et al.\ 1996ab; Smail
et al.\ 1997).

The ellipticities of the clusters as projected on the sky,
$\epsilon = 1-b/a$, were computed using all particles in the
the highest resolution
subgrids that were located within a distance of $r_{200}$ of the cluster
centers of mass.  
The probability,
$P(\epsilon)$, of observing a given projected
ellipticity for a given cluster was
computed by viewing each cluster from 500 random orientations and assembling
an appropriately normalized probability
distribution function.  Results
are shown in Fig.\ 7 for all 3 clusters at $\sigma_8 = 0.67$, 0.83, and
1.0.  Again, as with the evolution of the triaxiality parameter, each 
cluster exhibits its own particular evolution in projected shape.  Cluster
1 evolves toward being, on average, significantly 
flatter in projection at $\sigma_8 = 1.0$
than it was at $\sigma_8 = 0.67$.  Cluster 2, on the other hand, remains
approximately the same projected shape over the same time period and cluster 3
evolves toward being significantly rounder in projection on average.  
The median ellipticities for each of the clusters,
$\epsilon_{\rm med}$, are listed in Tables 1, 2, and 3.

\subsec{4.5 Substructure}

The redshift dependence of 
the fraction of observed clusters having a significant amount of
substructure is potentially a good indicator of the value of
density parameter.  This is due to the fact that in a universe in which 
$\Omega_0 \le 1$ density fluctuations cease to grow at redshifts of
order $\Omega_0^{-1} - 1$ and, so, moderate to low redshift clusters in
critical density universes are expected to contain substantial
amounts of substructure on average while in low density universes the
clusters should be much more regular. 
Wilson, Cole \& Frenk (1996) have
explored the possibility of using observations of weak lensing
by clusters to discriminate between universes with low and critical
values of $\Omega_0$ via a quantification of cluster substructure 
from the weak shear field.  Although initially optimistic,
the situation has become more murky recently with the 
realization that some of the simulated clusters used in the analysis
were inadequate.

The simulations discussed here are
restricted to a critical density universe and, hence, we do not
investigate the explicit $\Omega_0$ dependence of cluster
substructure with cosmological epoch
(this analysis will be performed in future simulations).
Rather, we have investigated the evolution of substructure in the clusters
from $\sigma_8 = 0.67$ to $\sigma_8 = 1.0$ in order to asses the 
degree to which substructure is erased over this period of time.

There are numerous methods by which cluster substructure can be
quantified but here we restrict the analysis to the
Dressler-Shectman $\Delta$ statistic (Dressler \& Shectman 1988).  
This choice is made
based on the results of Pinkey et al.\ (1996) who have subjected many
substructure tests to thorough analysis and conclude that by and large the 
Dressler-Shectman test tends to be the most sensitive to substructure.

The $\Delta$ statistic is defined by
$$\Delta = {1 \over N} \sum_{i=1}^N \delta_i, \;\;\;\;\;
  \delta_i^2 = {N_{\rm loc} \over \sigma^2} 
  \left[ \left( \overline{v}_i - \overline{v} \right)^2 +
  \left( \sigma_i - \sigma \right)^2 \right] \eqno (5) $$
where $N$ is the number of galaxies in the cluster and $\overline{v}_i$
and $\sigma_i$ are, respectively, the mean velocity and the velocity 
dispersion of the $N_{\rm loc}$ nearest neighbors to each galaxy.
The sensitivity of the $\Delta$ statistic is dependent upon the number
of neighboring galaxies used in the analysis and Bird (1995) finds the
test to be most sensitive for $N_{\rm loc} = \sqrt{N}$.  The statistic is
a measure of the correlation between the (projected) locations of
the galaxies in the cluster and their velocities.  In the case
of uncorrelated positions and velocities $\Delta \sim 1$.  The quantification
of substructure for a given cluster using only the computed value of $\Delta$
is insufficient, however, and in order to assess the likelihood of
real substructure within the cluster Monte Carlo simulations must be
performed.
Additionally, Crone, Evrard \& Richstone (1996) have pointed out that
since the value of $\Delta$ is not independent of $N$, the total number
of galaxies used in the analysis, in order to compare either observed or
theoretical clusters to one another it is necessary in the analyses
to select an 
identical number of galaxies for each cluster.

The simulations presented here are of insufficient resolution to
resolve the dark matter halos of individual galaxies (see \S 4.6 below)
and we instead investigate the substructure in the mass distribution.
(For all intents and purposes this is the nature of the substructure 
identified via weak lensing, though there is clearly good correspondence
between ``lumpiness'' in the mass and galaxy distributions in lensing
clusters.)  Substructure in the clusters was evaluated in the following
way.  Since the specific value of $\Delta$ for a given cluster
will depend on the angle from which it is viewed in projection, each
cluster was viewed from 500 random directions.  For each viewing angle, 
the projected locations of a randomly selected subset of the particles
was used to compute a value of $\Delta$.   Additionally, for each
viewing angle a Monte Carlo value of the statistic ($\Delta_{\rm rand}$)
was computed by randomly shuffling the velocities of the particles 
amongst their positions.  Mean values of $\Delta$ and $\Delta_{\rm rand}$,
along with their formal standard deviations, were then computed from 
the 500 individual values.  For each cluster a total of $N=1024$ particles
were randomly selected from all particles within the virial radius
($r_{200}$).  Different random sets of particles 
were used for each viewing angle and for the timesteps corresponding to
$\sigma_8 = 0.67$, 0.83, and 1.0.  Following Bird (1995), $N_{\rm loc}$
was taken to be $\sqrt{N} = 32$.

Tables 1, 2, and 3 list the mean values of the $\Delta$ statistic 
for each cluster computed from the 500 random viewing angles,
the mean values computed from the 500 Monte Carlo
position-velocity shuffles ($\Delta_{\rm rand}$), and the
corresponding 1-$\sigma$ errors.  Also listed are the ratios
 $\Delta / \Delta_{\rm rand}$, which indicate that all three clusters
contain significant amounts of substructure at each of the three epochs,
$\sigma_8$.  That is, within the virial radius substructure in the
mass distribution of the clusters
is not completely erased by the end of the simulation.  In the case of
cluster 2, the degree of substructure over and above the expectations
of random is roughly constant at a $\sim 3$-$\sigma$ level from
$\sigma_8 = 0.67$ to $\sigma_8 = 1.0$.  In the case of clusters 1 and
3, there is significantly less substructure at $\sigma_8 = 1.0$ than
at $\sigma_8 = 0.67$.  However, the ``erasure'' of substructure in these
two clusters over this time period is not quite monotonic.
Cluster 1 is a bit ``lumpier'' at $\sigma_8=0.83$ than it is at
either $\sigma_8 = 0.67$ or $\sigma_8 = 1.0$, while cluster 2 is a bit
smoother and less ``lumpy'' at $\sigma_8 = 0.83$ than it is at
either $\sigma_8 = 0.67$ or $\sigma_8 = 1.0$.

Bubble plots of the $\Delta$ test are very helpful for illustrating
visually both the location and amount substructure in a cluster.  Fig.\ 8
shows bubble plots for each of the clusters at the end of the
simulation.  The viewing angle for the projection of each cluster was
chosen  to be an angle for which the specific value of
$\Delta$ was identical to the mean value of the 500 random orientations.
The dots  in Fig.\ 8 show the spatial locations of the 1024 randomly selected
mass points used in the $\Delta$ statistic analysis. 
The circles,
all of which are centered on dots, 
have been scaled to have radii proportional to 
$\delta_i$ (see equation 5 above). 
The larger the circle, the larger is the local deviation of the
mean velocity and/or velocity dispersion from the global cluster value.
For clarity, circles are drawn
around only those mass points for which $\delta_i > 2\Delta$ (i.e.\
regions of most significant substructure).  Far from being smooth blobs
of mass, the clusters are all clearly ``lumpy'', each having of order 4 or
5 significant sub-lumps within the virial radius.

\subsec{4.6 Galaxy Halos}

It is often thought, erroneously, that purely dissipationless 
simulations are inadequate to study the dark matter halos  of galaxies
in a simulated cluster environment because all small dark matter 
concentrations are destroyed by purely numerical effects as they
orbit through the cluster.  However, Bromley et al.\ (1995)  have 
demonstrated that provided the force resolution is high galaxy-sized
dark matter halos will survive many orbits through the potential of a
large cluster and are not destroyed by purely numerical effects.

In the case of this work, the study of individual galaxy-sized dark
halos in the clusters is not a reasonable goal since the length scale
over which the force is non-Newtonian is somewhat too large, even in
the case of the highest resolution subgrids.  However, it can be seen
in Fig.\ 1 that there are numerous galaxy-sized concentrations of
dark matter in the outer regions of the
clusters and the clusters are not merely smooth
blobs of mass.  (The lumpiness of the mass distribution 
is borne out in part by the substructure analysis above.)  Due to the 
high particle density in the central regions and the choice of contrast
level, existing galaxy-sized concentrations in the inner regions
of the clusters are not visible in Fig.\ 1.

Although it is clearly an inadequate method for the generation of a
highly accurate catalog of galaxy-sized dark halos, a simple 
friends-of-friends algorithm was used to generate catalogs of groups
of particles in the highest resolution subgrids.  The groups were
selected to have overdensities
$\delta \go 1000$ (typical of the overdensity of the luminous region
of a bright galaxy) and masses $\go 10^{11}h^{-1} M_\odot$ (20 particles
or more).  
At the end of the simulations, $\sim 300$ such
objects were found within the Abell radii of clusters 1 and 3, and 
$\sim 200$ were found within the Abell radius of cluster 2.  However, 
owing to the large scale over which the force is softer than Newtonian,
the central $0.5h^{-1} {\rm Mpc}$ of each cluster is dominated by
a single huge ``halo'' of mass $\sim 10^{15} h^{-1} M_\odot$.

By nesting yet another subgrid within the highest resolution subgrid
(i.e.\ by performing a 4-level calculation), the effective length scale
over which the force is softer than Newtonian will be reduced significantly.
Within the Abell radius of a cluster it will then be possible to 
resolve confidently groups of particles that may be fairly associated 
with the dark
matter halos of individual galaxies and to
eliminate the artificial overmerging of halos in the central region.
Such analyses will be performed in future simulations.

\secno=5
\eqnum=0
\sec{5. DISCUSSION}

Using a hybrid N-body code in which high mass and spatial resolution
can be obtained in small regions of a very large total simulation
volume, the formation of three massive clusters was investigated.
The clusters were chosen to be typical of the most massive clusters
that would be present in a standard CDM universe at an epoch
corresponding to $\sigma_8 = 1.0$.  At highest resolution, the 
clusters consisted of $\sim 4\times 10^5$ particles within the Abell
radius at $\sigma_8 = 1.0$.
Although the clusters share
similar properties at the end of the simulation, the details of their
formation histories are quite different.

The properties which the clusters share are:
\vskip 0.2cm
\item{--} formation within the same large computational volume
($200^3 h^{-3} {\rm Mpc}^3$)
\item{--} masses of $M(r \le 1.5h^{-1} {\rm Mpc})
\sim 2\times 10^{15} h^{-1} M_\odot$ at the
end of the simulation
\item{--} similar spherically-averaged density profiles which are
well-fit by equation (2)
\item{--} similar values of the virial radius, $r_{200}$, and similar
values of the scale radius, $r_s$, at the end of the simulation
\item{--} accretion of $\sim 50$\% of the mass present at the end
of the simulation within the past 5 to 8 Gyr
\vskip 0.2cm

In terms of the details of the formation history of the clusters,
however, each cluster exhibits markedly individual behavior and
no single pattern 
dominates in the evolution of the following cluster properties:
\vskip 0.2cm
\item{--} mass accretion rate, $\dot{M}$
\item{--} scale radius, $r_s$
\item{--} three-dimensional shape (triaxiality parameter, $T$)
\item{--} two-dimensional (projected) ellipticity
\item{--} substructure (Dressler-Shectman $\Delta$ statistic)
\vskip 0.2cm

Based on the very small number of clusters presented here
it is difficult to make 
any statistically-sound conclusions about cluster formation and evolution.
However, the problems with standard
CDM notwithstanding, there
are certainly some interesting things to be noted.  The numerical clusters
are extremely massive and, so, correspond to rich clusters.
Rich clusters, being the brightest and most massive,
are likely to be those which will be studied observationally over the
widest range of redshifts and are the objects from which it is hoped
that cosmological constraints will arise.  Given the markedly different
formation histories of the clusters, a question raised by
this investigation is the degree to which observations of a sample
of rich clusters covering a wide range of redshifts can provide stringent
cosmological constraints.

Although the numerical clusters studied here do not constitute a statistically
large sample, it is clear that such large clusters do not form a simple
1-parameter family as far as their evolution history is concerned, even though
the clusters have similar masses and spherically-averaged density
profiles at the present day.  The use of observations of cluster evolution
to constrain cosmological models may yet be viable but the cautious suggestion
from this work is that this may not be completely straightforward.
Considerably more work on the details of the formation history of clusters
in various models of structure formation is necessary in order to determine
both the degree to which cosmological conclusions can be drawn from
observations of clusters at different epochs and also the requisite
size of a sample of observed clusters which would insure those cosmological
conclusions to be statistically reliable.

\sec{ACKNOWLEDGMENTS}

Generous amounts of CPU time on the Cray J-90 at the Max-Planck-Institut
f\"{u}r Astrophysik and enlightening conversations with Simon White
are gratefully acknowledged.  This work was supported in part by
the NSF under contract AST-9616968.

\sec{REFERENCES}
\secno=0
\eqnum=0

\noindent 
Bahcall, N. A. \& Cen, R. 1993, ApJ, 407, L49

\noindent
Bartelmann, M., Ehlers, J. \& Schneider, P. 1993, A\&A, 280, 351

\noindent
Beers, T. C. \& Geller, M. J. 1983, ApJ, 274, 491

\noindent
Bird, C. M. 1994a, ApJ, 422, 480

\noindent
Bird, C. M. 1994b AJ, 107, 1637

\noindent
Bird, C. M. 1995, ApJ, 445, L81

\noindent
Bonnet, H., Mellier, Y., \& Fort, B., 1994, ApJ, 427, L83

\noindent
Bower, R. G. \& Smail, I. 1997, MNRAS, in press (astro-ph/9612151)

\noindent
Bromley, B. C., Warren, M. S., Zurek, W. H., \& Quinn, P. J. 1995,
AIP Conference 

Proceedings 336, 433

\noindent
Bunn, E. \& White, M. 1997, ApJ, 480, 6

\noindent
Carlberg, R. G. 1994, ApJ, 433, 468

\noindent
Cole, S. M. \& Lacey, C. G. 1996, MNRAS, 281, 716

\noindent
Crone, M. M., Evrard, A. E. \& Richstone, D. O. 1996, ApJ, 467, 489

\noindent
Davis, D. S. \& Mushotzky, R. F. 1993, AJ, 105, 409

\noindent
Deltorn, J.-M., Le Fevre, O., Crampton, D., \& Dickinson, M. 1997,
ApJ (Letters), in press 

(astro-ph/9704086)

\noindent
Dressler, A. \& Shectman, S. A. 1988, AJ, 95, 985

\noindent
Dubinski, J. \& Carlberg, R. 1991, ApJ, 378, 496

\noindent
Eke, V. R., Cole, S. \& Frenk, C. S. 1996, MNRAS, 282, 263

\noindent
Fahlman, G., Kaiser, N., Squires, G., \& Woods,
D., 1994, ApJ, 437, 56

\noindent
Jones, C. \& Forman, W. 1984, ApJ, 276, 38

\noindent
Kauffmann, G. \& White, S. D. M. 1993, MNRAS, 261, 921

\noindent
Kneib, J-P, Ellis, R. S., Smail, I., Couch, W. J.,
 \& Sharples, R. M.
1996, ApJ, 471, 643

\noindent
Lacey, C. \& Cole, S. 1993, MNRAS, 262, 627

\noindent
Luppino, G. \& Kaiser, N. 1997, ApJ, 475, 20

\noindent
Miyaji, T., Mushotzky, M., Lowenstein, M., Serlemitsos, P. J,
Marshall, F. E., Petre, R., Jahoda, K. M., Boldt, E. A., 
Holt, S. S., Swank, J., Szymkowiak, A. E., \& R. Kelley

1993, ApJ, 419, 66

\noindent
Mohr, J. J., Evrard, A. E., Fabricant, D. G., \& Geller, M. J. 1995,
ApJ, 447, 8

\noindent
Mushotzky, R. 1993, ANYAS, 699, 184

\noindent
Navarro, J. F., Frenk, C. S. \& White, S. D. M. 1995, MNRAS, 275, 720

\noindent
Navarro, J. F., Frenk, C. S. \& White, S. D. M. 1996a, ApJ, 462, 563

\noindent
Navarro, J. F., Frenk, C. S. \& White, S. D. M. 1996b, ApJ, submitted
(astro-ph/9611107)

\noindent
Pinkey, J., Roettiger, K., Burns, J. O., \& Bird, C. M. 1996, ApJS,
104, 1

\noindent 
Richstone, D., Loeb, A. \& Turner, E. L. 1992, ApJ, 393, 477

\noindent
Seitz, C., Kneib, J.-P., Schneider, P., \& Seitz, S.,
1996, A\&A, 314, 707

\noindent
Smail, I. \& Dickinson, M. 1995, ApJ, 455, L99

\noindent
Smail, I., Ellis, R. S., Fitchett, M. J., \& Edge,
A. C., 1995, MNRAS, 273, 277

\noindent
Smail, I., Ellis, R. S., Dressler, A., Couch, W. J.,
Oemler, A., Sharples, R., \& Butcher, H.

1997, ApJ, 479, 70

\noindent
Squires, G., Kaiser, N., Babul, A., Fahlman, G., Woods, D.,
Neumann, D. M., \& B\"ohringer, 

H. 1996a, ApJ, 461, 572

\noindent
Squires, G., Kaiser, N., Fahlman, G., Babul, A.,
et al., 1996b, ApJ, 469, 73

\noindent
Steidel, C., Dickinson, M., Meyer, D., Adelberger, K. \& Semback, K.
1997, ApJ, submitted 

(astro-ph/9610230)

\noindent
Tormen, G., Bouchet, F. \& White, S. D. M. 1997, MNRAS, 286, 865

\noindent
West, M. J. \& Bothun, G. D. 1990, ApJ, 350, 36

\noindent
White, S. D. M., Briel, U. G. \& Henry, J. P. 1993, MNRAS, 261, L8

\noindent
White, S. D. M., Efstathiou, G. \& Frenk, C. S. 1993, MNRAS, 262, 1023

\noindent
Wilson, G., Cole, S. \& Frenk, C. S. 1996, MNRAS, 282, 501

\noindent
Viana, T. P. \& Liddle, A. R. 1996, MNRAS, 281, 323

\noindent
Villumsen, J. V. 1989, ApJS, 71, 407

\noindent
Zabludoff, A. I. \& Zaritsky, D. 1995, ApJ, 447, L21

\sec{FIGURE CAPTIONS}
\parindent=0pt

Fig.\ 1a:  Grey-scale images of cluster 1 at $\sigma_8 = 1.0$.  The top
panel shows the logarithm of the
mass density along the line of sight in the 
high-resolution subgrid (subgrid \#2) and the center panel shows the same
for the low resolution subgrid (subgrid \#1).  The pixel sizes in the
image are equal to the grid cell sizes in the subgrids and the projection
is of a $8.3 h^{-1} {\rm Mpc} \times 8.3h^{-1} {\rm Mpc} \times 
8.3h^{-1} {\rm Mpc}$ cube, centered on the center of mass of the cluster.
The bottom panel shows a comparison of the line of sight mass density
in the two subgrids, where the grey scale indicates the logarithm of
the mass density in subgrid \#2 divided by the mass density in 
subgrid \#1, computed at the (low) resolution of subgrid \#1.  Overall,
the mass density in the two subgrids compares well (see text).

\vskip 0.5cm
Fig.\ 1b: Same as Fig.\ 1a, but for cluster 2.

\vskip 0.5cm
Fig.\ 1c: Same as Fig.\ 1a, but for cluster 3.

\vskip 0.5cm
\noindent
Fig.\ 2:  The infall distance of particles into the clusters.  Solid points
with error bars show the mean distance streamed since the beginning
of the simulation as a function of the final location of the particles,
computed relative to the cluster centers of mass.  Error bars indicate
one standard deviation.  The open squares show the maximum initial distance
of any one particle from the center of mass of its cluster as a function
of its location at the the end of the simulation.  

\vskip 0.5cm
\noindent
Fig.\ 3:  Mass of the clusters contained within the Abell radius as
a function of lookback time.  The contained mass has been normalized
by the contained mass at the end of the simulation.  All three clusters
have masses $\sim 2 \times 10^{15} h^{-1} {\rm Mpc}$ within the Abell
radius at the end of the simulation (see Tables 1, 2, and 3).

\vskip 0.5cm
\noindent
Fig.\ 4: Rate at which mass is accreted, $\dot{M}$, within the Abell
radius as a function of lookback time.  The accretion rate is normalized
by the mean rate, $\left< \dot{M} \right>$, at which mass is accreted
within the Abell radius
between lookback times corresponding to $z=2$ and $z=0$.

\vskip 0.5cm
\noindent
Fig.\ 5: Spherically averaged density profiles for each of the clusters,
evaluated at $\sigma_8 = 1.0$.  Open squares indicate $\rho(r)$ computed
using the high-resolution subgrid particles; filled triangles indicate
$\rho(r)$ computed using the low-resolution subgrid particles.

\vskip 0.5cm
\noindent
Fig.\ 6:  Cluster overdensities as a function of radius (scaled by
$r_{200}$, the virial radius) for $\sigma_8 = 0.67$, 0.83 and
1.0.  Results for cluster 1 are shown in the three panels on the left,
results for cluster 2 are shown in the three central panels, and results
for cluster 3 are shown in the three panels on the right.  The solid
line indicates the best-fit density profile of the form of equation
(2), with the corresponding value of the scale radius, $r_s$, indicated
in each of the individual panels of the figure.

\vskip 0.5cm
\noindent
Fig.\ 7:  Projected ellipticity distributions for each of the clusters
as a function of $\sigma_8$.  The dashed line indicates $\sigma_8 = 0.67$,
the dotted line indicates $\sigma_8 = 0.83$, and the solid line
indicates $\sigma_8 = 1.0$.

\vskip 0.5cm
\noindent
Fig.\ 8:  Bubble plots of the $\Delta$ test for each cluster at
$\sigma_8 = 1.0$.  The linear scale of the projection is
$5h^{-1} {\rm Mpc}$ by $5 h^{-1} {\rm Mpc}$.  Dots indicate the
spatial location of the mass points used in the evaluation of
$\Delta$ and the circles have radii proportional to $\delta_i$
(see text).  The degree of substructure apparent in these projections
is indicative of the
mean value, based on the results of 500 random viewing angles.

\vfil\eject
\topinsert
\vskip -1.0in
{\centerline{\epsfxsize=6.4in \epsfbox{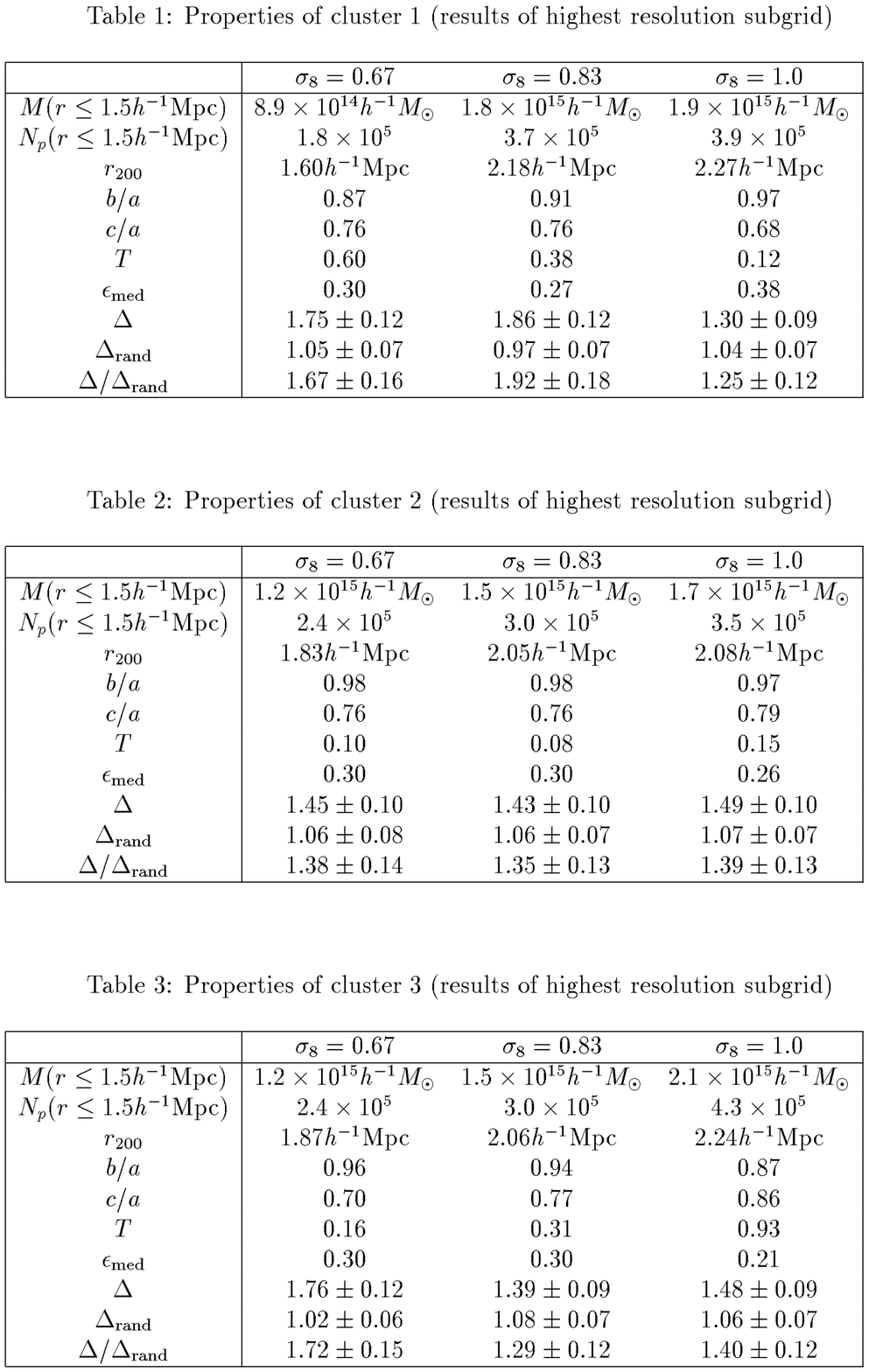}}}
\endinsert

\vfil\eject
\topinsert
\vskip -1.0in
{\centerline{\epsfxsize=7.1in \epsfbox{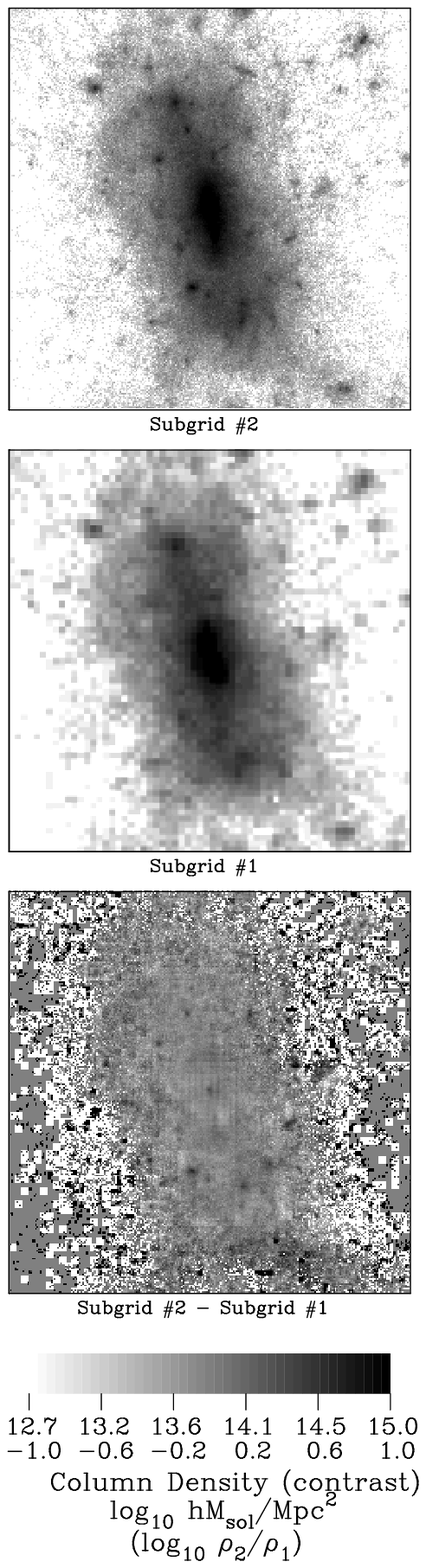}}
\vskip -1.3in
{\centerline {\bf Figure 1a}}}
\endinsert

\vfil\eject
\topinsert
\vskip -1.0in
{\centerline{\epsfxsize=7.1in \epsfbox{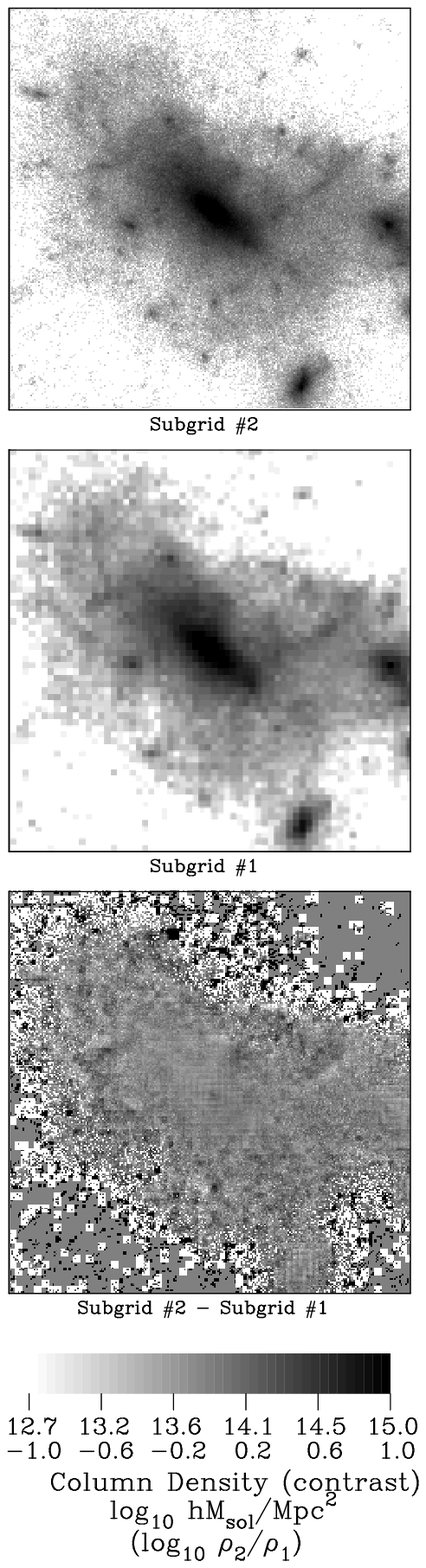}}
\vskip -1.3in
{\centerline {\bf Figure 1b}}}
\endinsert

\vfil\eject
\topinsert
\vskip -1.0in
{\centerline{\epsfxsize=7.1in \epsfbox{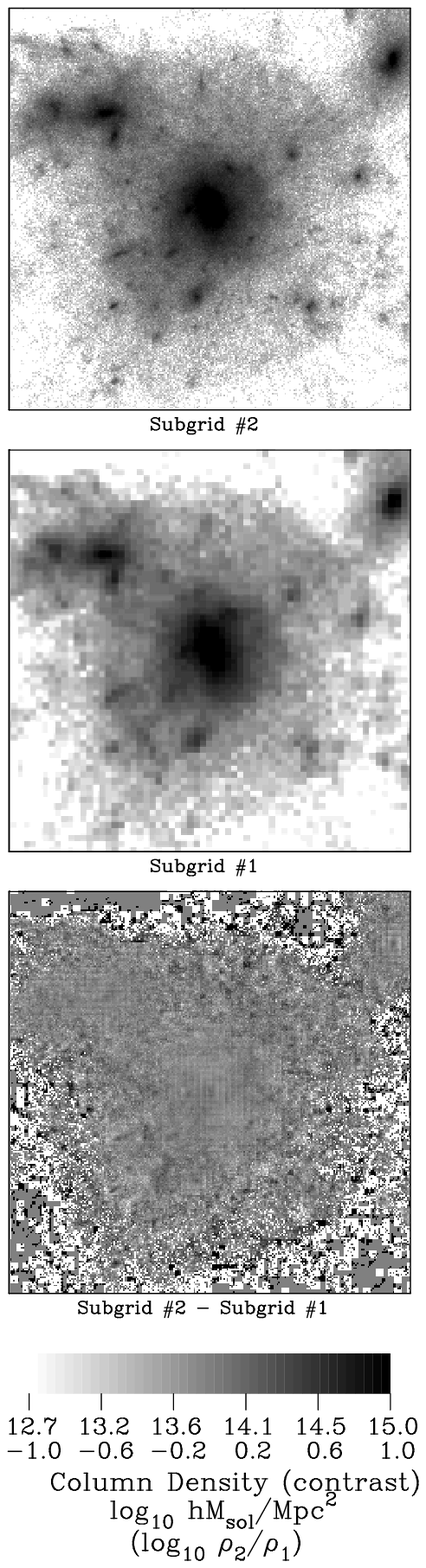}}
\vskip -1.3in
{\centerline {\bf Figure 1c}}}
\endinsert

\vfil\eject
\topinsert
\vskip -1.0in
{\centerline{\epsfxsize=6.2in \epsfbox{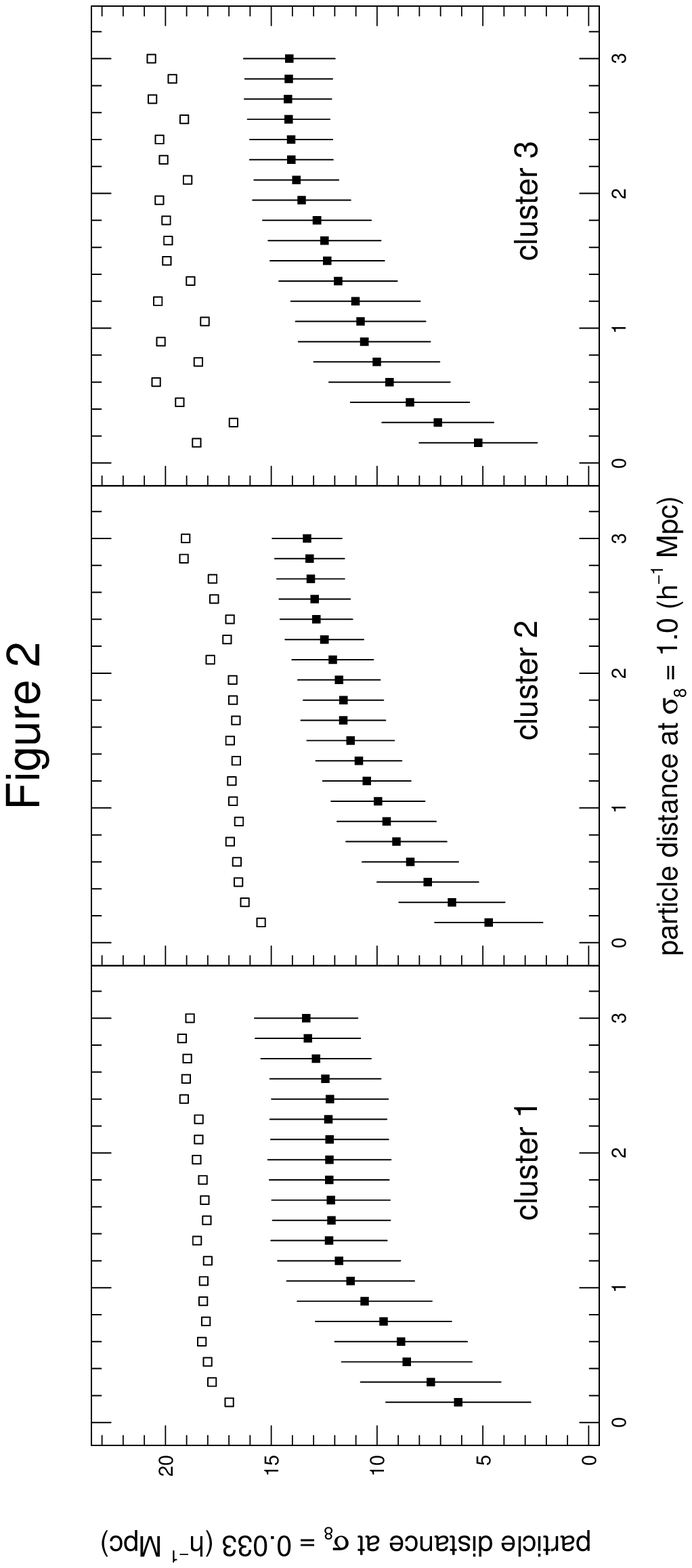}}}
\endinsert

\vfil\eject
\topinsert
\vskip -2.4in
{\centerline{\epsfxsize=6.2in \epsfbox{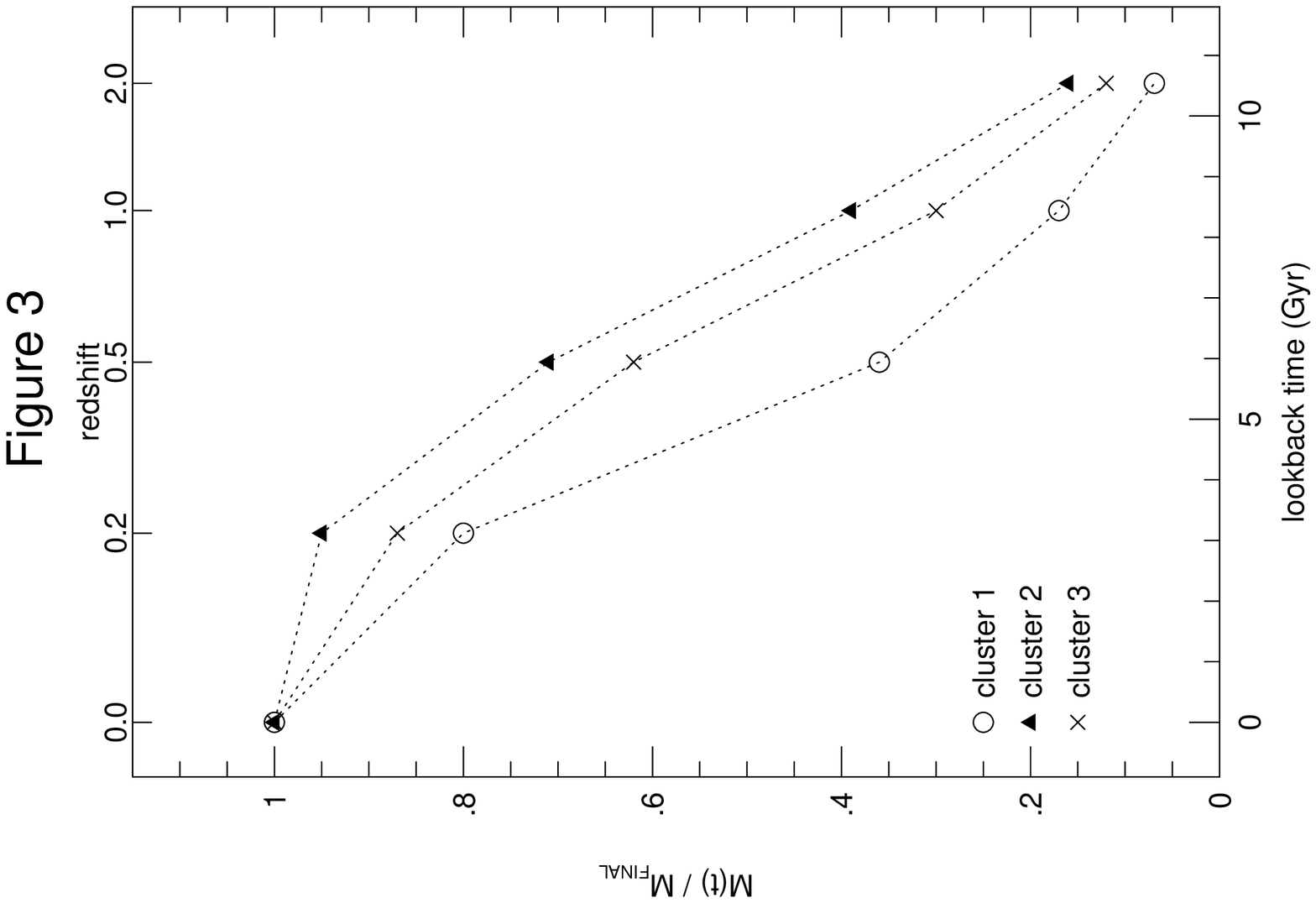}}}
\endinsert

\vfil\eject
\topinsert
\vskip -1.0in
{\centerline{\epsfxsize=6.2in \epsfbox{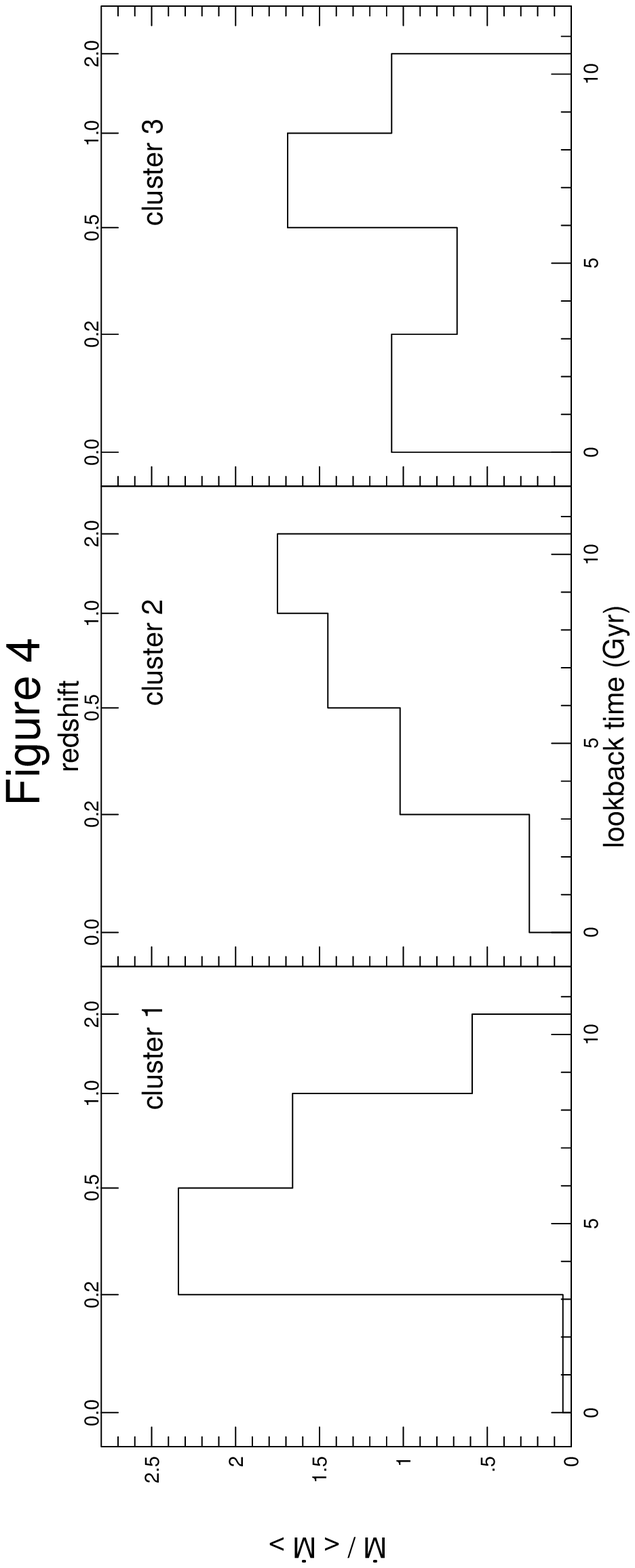}}}
\endinsert

\vfil\eject
\topinsert
\vskip -1.0in
{\centerline{\epsfxsize=6.2in \epsfbox{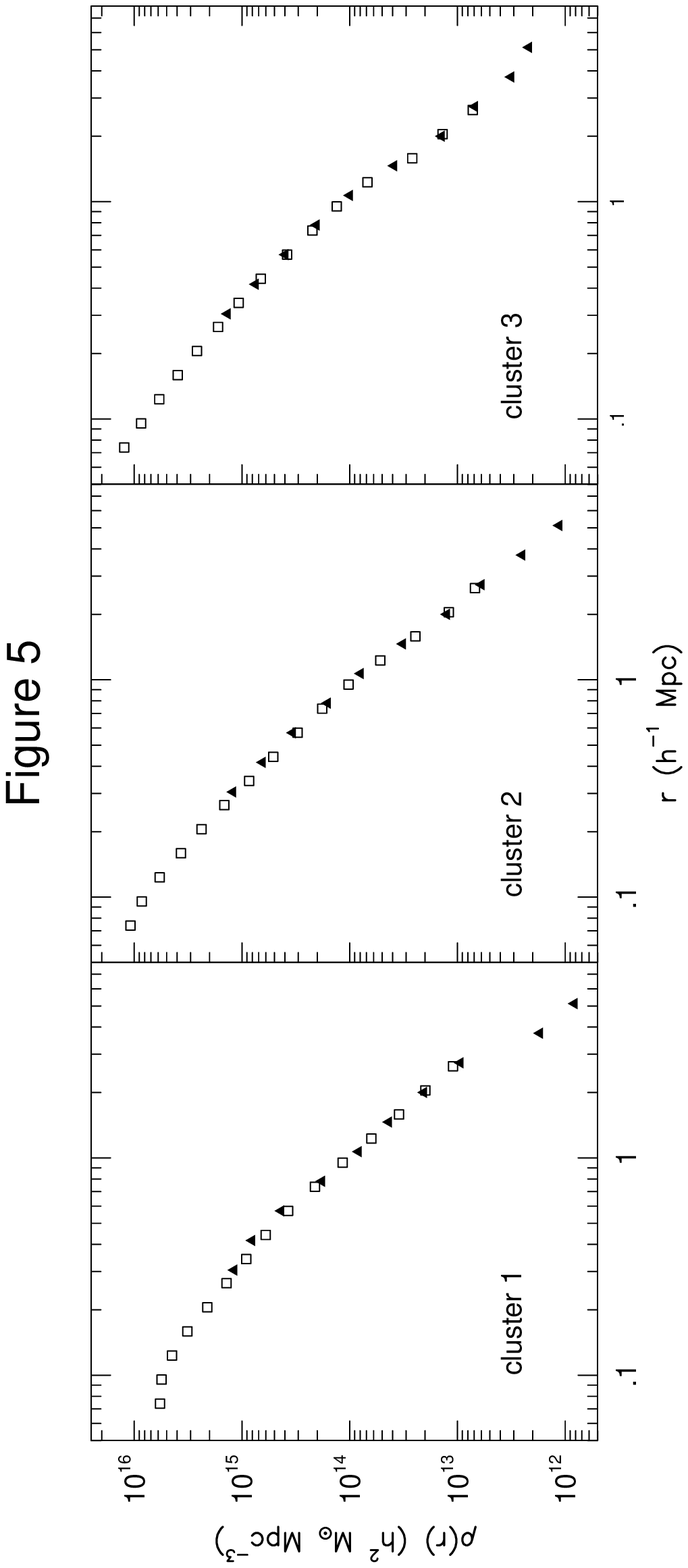}}}
\endinsert

\vfil\eject
\topinsert
\vskip -1.0in
{\centerline{\epsfxsize=6.2in \epsfbox{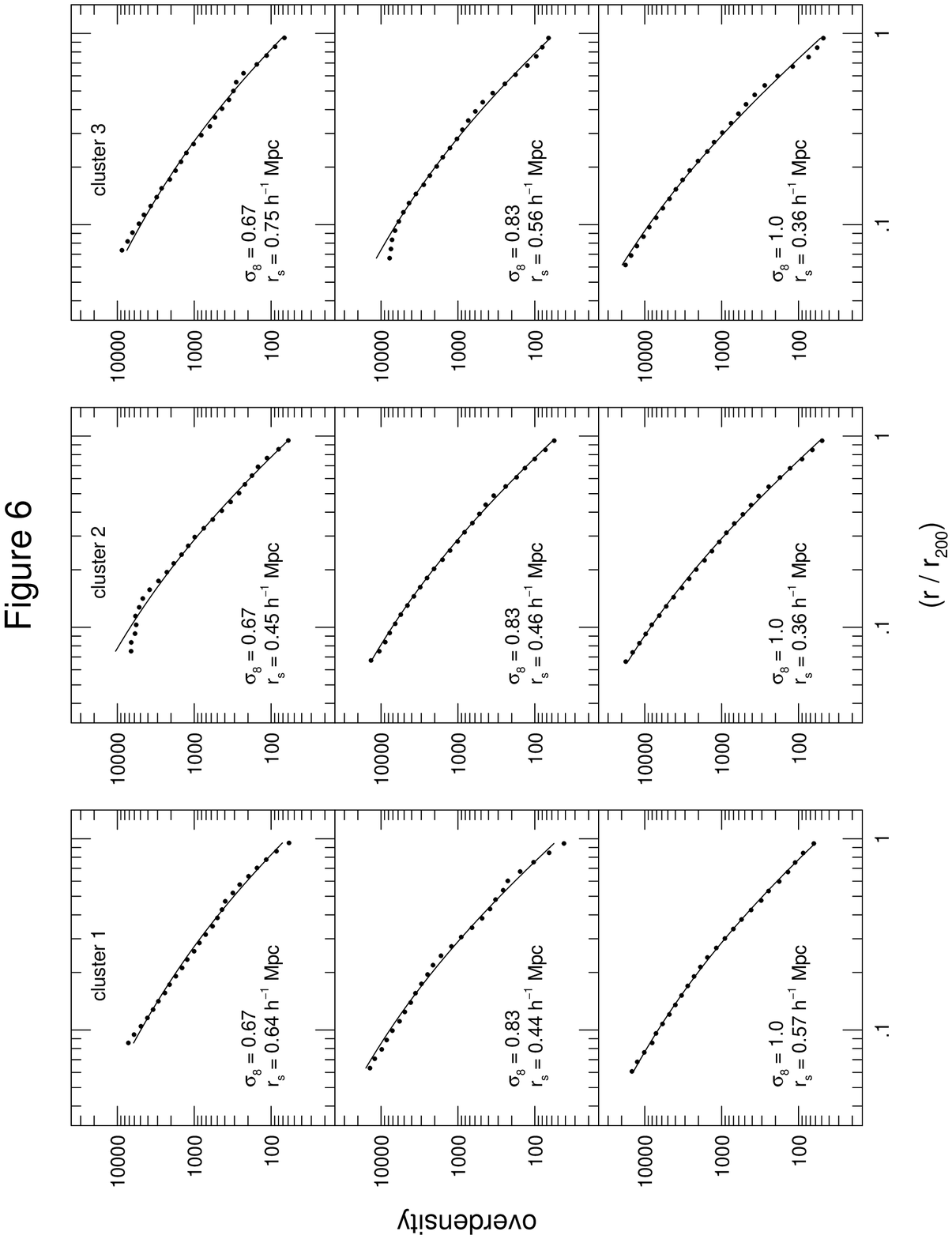}}}
\endinsert

\vfil\eject
\topinsert
\vskip -1.0in
{\centerline{\epsfxsize=6.2in \epsfbox{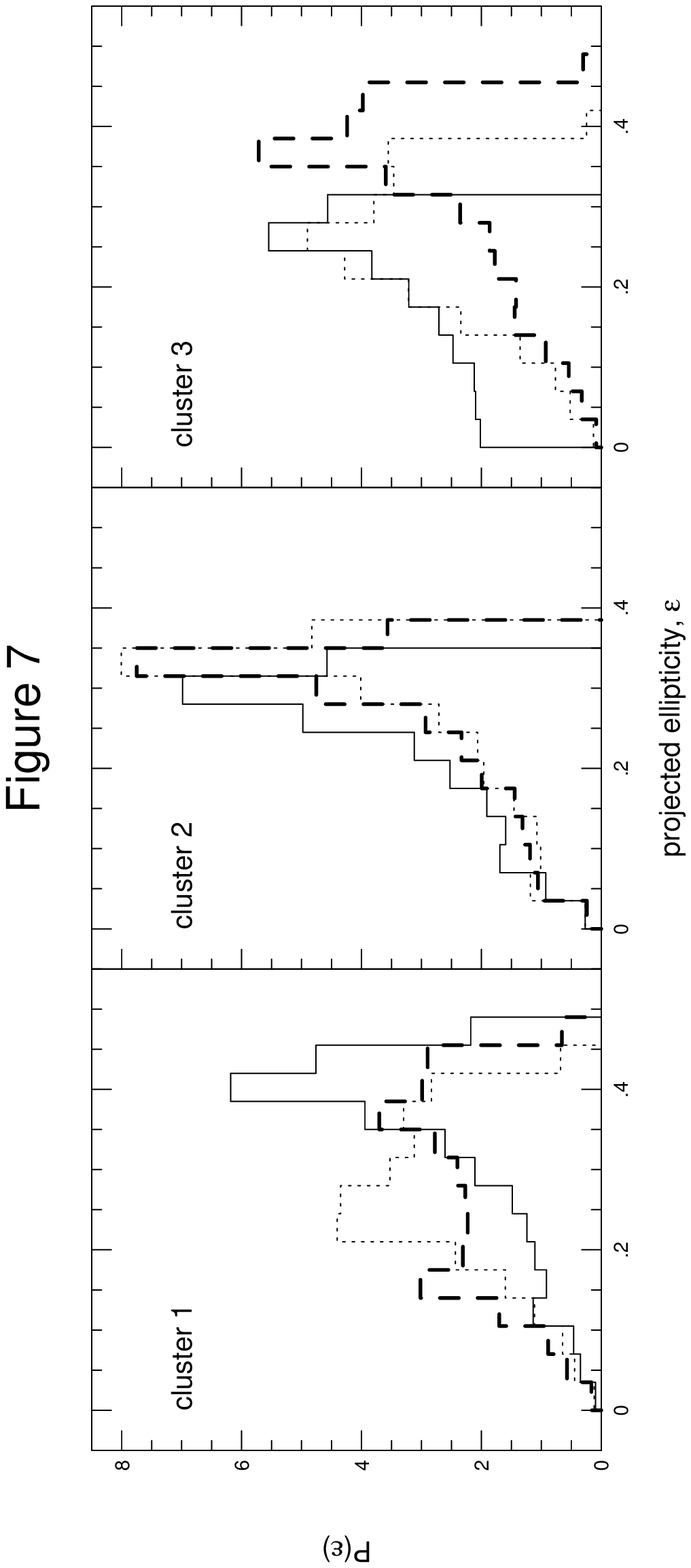}}}
\endinsert

\vfil\eject
\topinsert
\vskip -1.0in
{\centerline{\epsfxsize=6.2in \epsfbox{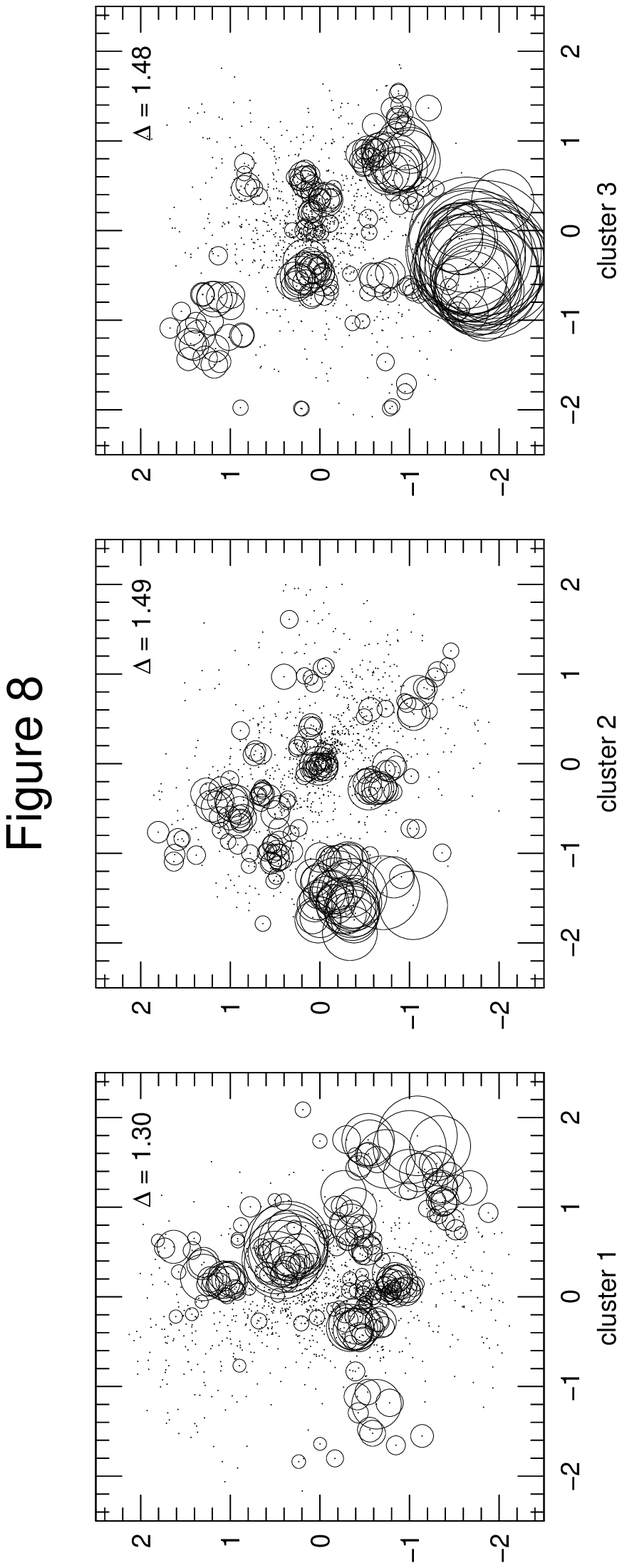}}}
\endinsert

\bye